\documentclass{elsart}

\usepackage{graphicx}

\usepackage{amsmath,amssymb}
\usepackage[english]{babel}

\graphicspath{{figures/}}

\newcommand{\add}[1]{#1}
\newcommand{\stdPictureWidth}{0.6\textwidth}
\newcommand{\widePictureWidth}{0.6\textwidth}

\newcommand{\Neel}{N\'eel}
\newcommand{\MB}{CuB$_2$O$_4$}

\begin{document}

\begin{frontmatter}

\title{Magnetic phase diagram copper metaborate \MB\ in magnetic field parallel c-axis: resonant, magnetic and magnetoelastic investigations}

\author[this]{A.~Pankrats}
\author[this]{G.~Petrakovskii}
\author[this]{V.~Tugarinov}
\ead{vit@iph.krasn.ru}
\author[this]{K.~Sablina}
\author[this]{L.~Bezmaternykh}
\author[poland]{L.~Szymczak}
\author[poland]{M.~Baran}
\author[poland]{B.~Kundys}
\author[poland]{A.~Nabialek}
\address[this]{Institute of Physics, SB RAS, 660036 Krasnoyarsk, Russia}
\address[poland]{Institute of Physics PAS, Warszawa 02-668, Poland}

\begin{abstract}
The magnetic phase diagram in a single crystal of copper metaborate \MB\ in a magnetic field parallel to a tetragonal axis $c$ has been investigated.  From the resonant, magnetic and magnetostrictive data the phase diagram of \MB\ on a plane ``temperature - magnetic field'' is constructed. The magnetic incommensurate-commensurate phase transition is caused by the saturation of weak subsystem of copper ions in the strong magnetic field $H\|c$.
\end{abstract}


\end{frontmatter}


\section{Introduction}

The tetragonal crystal of copper metaborate \MB\ has the complex magnetic structure which examination was carried out by various experimental methods including a neutron scattering~\cite{Roessli,Boehm}, $\mu$SR~\cite{Boehm_mSR}, a magnetic resonance~\cite{Pankrats_2000} and magnetic measurements~\cite{Balaev_2000}. Neutron investigations showed that at ${T<T_{spon}=9.5}$~K the magnetic state of a crystal is incommensurate with a magnetic propagation vector directional along a tetragonal axis. Resonant and magnetic measurements~\cite{Pankrats_2003} allowed to assume that in the temperature interval from $9.5$~K up to \Neel\ temperature $T_N=20$~K the ground state is also modulated and long periodical. In a magnetic field perpendicular to the tetragonal axis both modulated states transform into the field-induced weak ferromagnetic state with the magnetic moment laying in a basal plane of a crystal. 

The magnetic phase diagram of copper metaborate in the perpendicular field is given in~\cite{Pankrats_2003}. The purpose of the work is the experimental investigation of the magnetic phase diagram of \MB\ in the magnetic field along a tetragonal axis. Phase transitions are explored by experimental methods sensitive to magnetic state of the crystal: electron-spin resonance~(ESR), magnetic and magnetostrictive measurements.
\section{Experimental results}

\add{Resonant and magnetic measurements are performed on single crystals of \MB\ growthing by spontaneous crystallization method. The samples having a form of a plate with sizes of up to ${2 \times 7 \times 7}$~mm$^3$ which were cut out in $(100)$ and $(110)$ crystallographic planes were used in magnetostriction measurements.
Resonant investigations are carried out on computer-controlled magnetic resonance spectrometer with pulsed magnetic field~\cite{Tugarinov_2004}. Magnetostriction measurements were done using the capacitance technique. Magnetic measurements are performed by a SQUID magnetometer MPMS-5.}

\begin{figure}[htbp]
	\centering
		\includegraphics[width=0.8\textwidth]{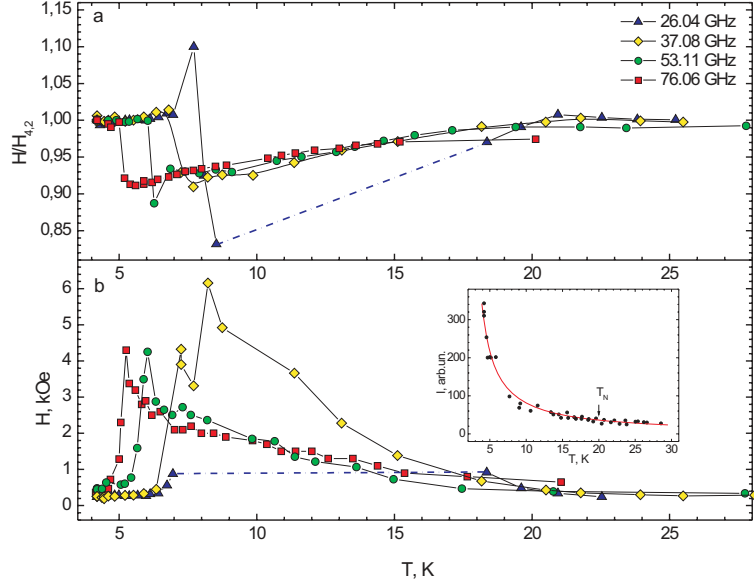}
	\caption{Temperature dependencies of resonance field~(a) and line width~(b) at $H\|$c and various frequencies. Inset: Temperature dependence of the intensity of magnetic resonance at the frequency $41.4$~GHz.}
	\label{fig:resonance}
\end{figure}

\begin{figure}[htbp]
	\centering
		\includegraphics[width=\widePictureWidth]{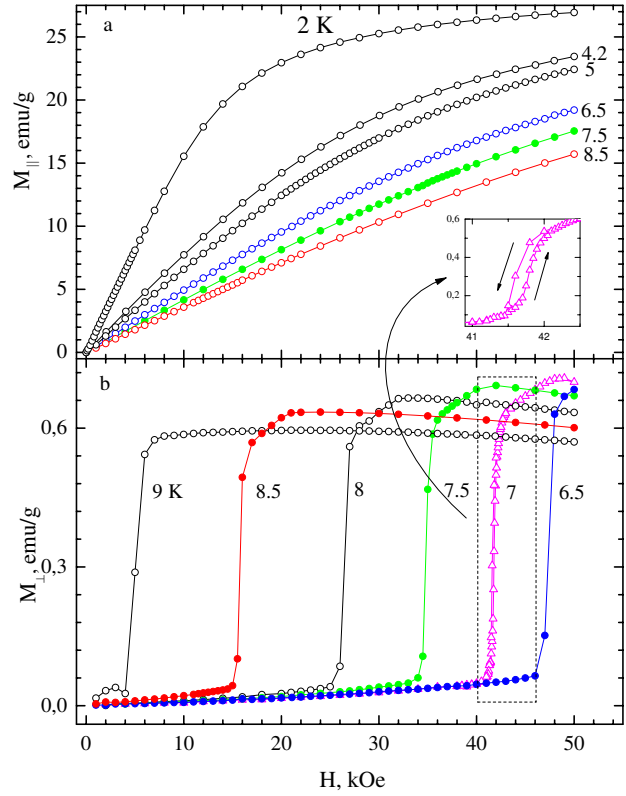}
	\caption{Field dependencies of longitudinal~(a) and transversal~(b) magnetization measured at various temperatures and $H\|c$.}
	\label{fig:magnetic}
\end{figure}

\begin{figure}[htbp]
	\centering
		\includegraphics[width=\stdPictureWidth,trim=0 0 0 0,clip]{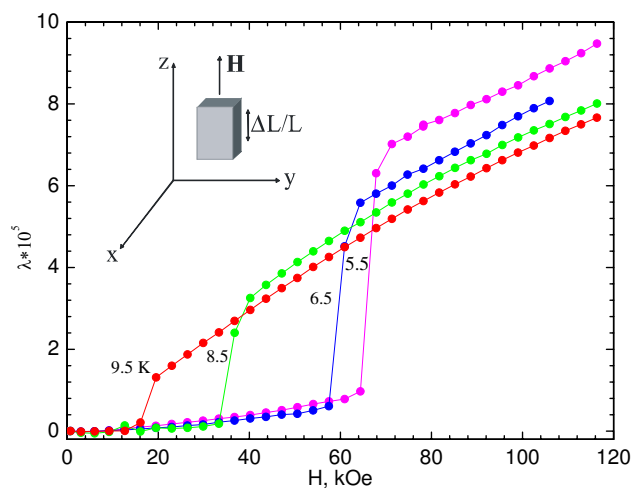}
	\caption{Field dependencies of longitudinal magnetostriction at various temperatures and $H\|c$}
	\label{fig:elastic}
\end{figure}

Temperature dependencies of resonance field and line width  measured for $H\|c$ at different frequencies have sharp anomalies~(Fig.\ref{fig:resonance}). The maximal line broadening decreases with the increasing of a frequency and, correspondingly, a magnetic field.
To ensure that anomalies are due to the transition from incommensurate to commensurate weak ferromagnetic state the dependencies of longitudinal and transversal magnetization have been measured at $H\|c$. The longitudinal magnetization~(Fig.~\ref{fig:magnetic}a) increases smoothly with a field increase at any temperatures below \Neel\ temperature, however, the dependences become more nonlinear at lower temperatures. The base saturation level of a longitudinal magnetization is reached at $T=2$~K in a field about $30$~kOe after which, as show measurements up to $350$~kOe, the weak linear rise of a magnetization is observed. At the same time the transversal magnetization~(Fig.~\ref{fig:magnetic}b) is near zero below some critical field and have a jump about $0.6$~emu/g at the critical field which is increasing with temperature lowering. The magnetization in magnetic field higher than critical one increases first reaching some maximum and then decreases gradually with further increase in a field.
The field dependences of transversal magnetization are also measured at ${T>T_{spon}}$ and differ considerably from the low temperature ones. The magnetization at ${T>T_{spon}}$ rises continuously with magnetic field showing a kink point at the critical field. \add{In this temperature interval the critical values of fields are much lower than at $T<9.5$~K.}
\begin{figure}[htbp]
	\centering
		\includegraphics[width=\stdPictureWidth]{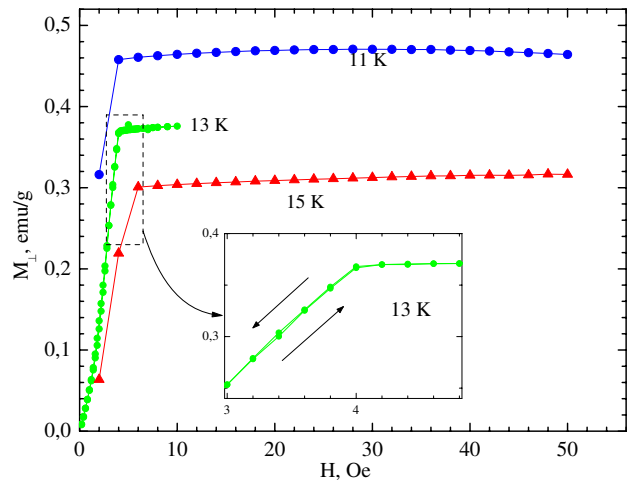}
	\caption{\add{Field dependencies of transversal magnetization measured at various temperatures above $9.5$~K and $H\|c$.}}
	\label{fig:magnetic_above_10K}
\end{figure}
Magnetostrictive examinations also allow to register transition between phases. We measured field dependences of a longitudinal and transversal magnetostriction for $H\|c$ at various temperatures. The typical field dependences of a longitudinal magnetostriction at temperatures below $9.5$~K are shown in Fig.~\ref{fig:elastic}. All dependences have common character: weak rise of a magnetostriction above and below a critical field and the jump at the transition. The field dependences of a transversal magnetostriction have a similar view.
\section{Discussion}
Complexity of the magnetic phase diagram of copper metaborate is due to coexistence of two subsystems of copper ions with a various degree of the magnetic order and different magnetic dimensionality. 
It is known~\cite{Boehm,JETP_2001}, that in unit cell of \MB\ 12 copper ions $Cu^{2+}$ occupy two non-equivalent positions --- $4b$ and $8d$. Four ions in a position $4b$ form the three-dimensional subsystem that is magnetically ordered below \Neel\ temperature~${T_N=20}$~K (the strong subsystem A). Other eight ions form a weak ordered subsystem B which is the one-dimensional and is partially polarized due to an exchange interaction with ions of the strong subsystem.

\begin{figure}[htbp]
	\centering
		\includegraphics[width=\stdPictureWidth]{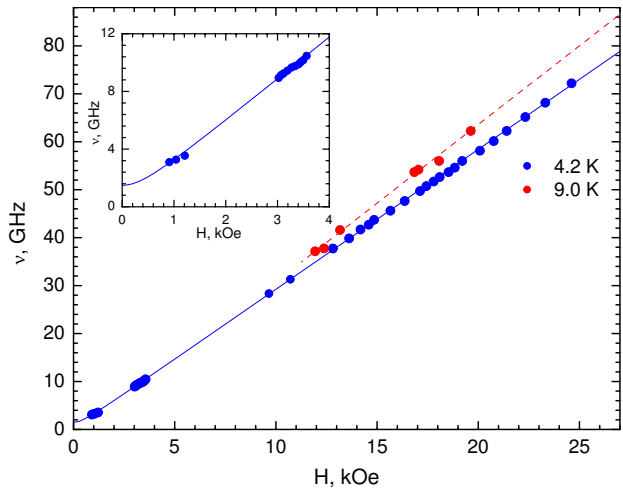}
	\caption{Frequency-field dependence of \MB\ at $4.2$~K (spiral phase) and $9.0$~K (weak ferromagnetic phase)}
	\label{fig:metaborate_freq-field_dependence}
\end{figure}

The analysis of resonant properties of \MB\ allow to assume that the resonant absorption in this crystal at orientation of a magnetic field along a tetragonal axis is connected with the weak subsystem B. The following arguments confirm such explanation. First, below \Neel\ temperature a strong subsystem  at $H\|c$ in both commensurate and incommensurate states can be considered as easy-plane antiferromagnet. The spectrum of antiferromagnetic resonance (AFMR) of this subsystem at such orientation of a field contains a branch with nonlinear frequency-field dependence and an energy gap $\omega_c\approx\gamma\sqrt{2 H_E H_A}$, where $H_E$ and $H_A$ --- effective fields, accordingly, of exchange and anisotropy with respect to the tetragonal axis (the second branch of AFMR is Goldstone with $\omega=0$). As the field of anisotropy $H_A$ for \MB\ is unknown it is impossible to estimate the value of the gap but the usual its value in uniaxial antiferromagnets is about several hundreds in GHz. At the same time the magnetic resonance data at this field orientation~(Fig.~\ref{fig:metaborate_freq-field_dependence}) show that the frequency-field dependences are near-linear with neglible values of the gaps in both  commensurate and incommensurate states.
The data of inelastic neutron scattering~\cite{JMMM_2002,JMMM_2004} also show two spin wave branches, one of which, high-energy, has a gap $840$~GHz at ${T=1.5}$~K and is attributed to the strong subsystem A, and another branch with small initial splitting is referred by authors to a weak subsystem.

It is visible from the inset of ~Fig.\ref{fig:resonance} that the temperature dependence of line intensity is well described as $I\sim C/(T-\theta)$ with ${\theta\approx2}$~K what is usual for the disordered systems. 
At last, the full absence of anomalies of intensity and line width at the $T_N$ of the strong subsystem also allows to connect the observed resonant absorption to a weak subsystem B.

In our opinion the broadening of the resonance line near the temperature of phase transition is caused by the fluctuations that are increasing with the approach to the transition. This result correlates with anomalous increase of diffuse neutron scattering  at phase transition~\cite{Roessli}. \add{With increase of frequency and, accordingly, a resonant field the external magnetic field suppresses fluctuations more strongly, and line spreading decreases.}

\add{Field dependences of longitudinal magnetization copper metaborate, measured along a tetragonal axis, confirm the assumption about a various degree of the magnetic order in subsystems A and B.}
Calculations show that saturation magnetizations of subsystems  along the $c$-axis are: ${M^A_S=13.54}$~emu/g and ${M^B_S=27.07}$~emu/g. The field dependence of longitudinal magnetization measured at $T=2$~K have two areas: sharp nonlinear increasing with saturation in fields approximately up to 30 kOe and practically linear subsequent increase in fields up to 350 kOe. Saturation occurs at the level corresponding to a weak subsystem, hence, this subsystem is saturated in magnetic fields up to $30$~kOe at $T=2$~K. The subsequent linear rise of magnetization is caused, mainly, by the antiferromagnetic susceptibility of the strong subsystem A. 
 
It is obvious that the jumps of magnetization in a basal plane at ${T<9.5}$~K (Fig.~\ref{fig:magnetic}b) are caused by the phase transition into the field-induced weak ferromagnetic state. The inset in Fig.~\ref{fig:magnetic} shows the magnetic hysteresis, hence at ${T<9.5}$~K the phase transition at $H\|c$, as well as in basal plane~\cite{ApplPhA_2002} is of the first order. There is no magnetic hysteresis of transversal magnetization at the critical field at ${T>T_{spon}}$ \add{(Fig.~\ref{fig:magnetic_above_10K})}. Thus the phase transition has the second order at the temperature range from $9.5$ to $20$~K. 

\add{The initial part of the field dependences of transversal magnetization also are partially caused by magnetization due to the projection of a field in the basic plane, and also nonideal orthogonality of measuring coils and the magnetic field.}
The measured transversal magnetization is the sum of contributions of both magnetic subsystems A and B, therefore the field dependence of the magnetization in the basal plane above the critical field is defined by two processes. On the one hand, the saturation of the weak subsystem along the $c$-axis in strong fields results in reduction of its contribution to total magnetization in the basal plane. On the other hand, there is an increase in the total magnetization caused by the increasing of a magnetic field component in the basal plane due to the nonideal its orientation along the $c$-axis. Clearly, that the main contribution to the last process is caused by the strong subsystem. Due to the competition of these two processes the total magnetization  increases first with the increasing of a magnetic field above its critical value, then starts to fall.

The analysis of the results on magnetostriction shows that its smooth increase with a field above and below the critical value is caused by the contribution of magnetized weak-ordered subsystem B. Taking into account that both longitudinal and transversal magnetostrictions have similar field dependences we can assume that the magnetostriction is a volume dilatation at $H\|c$. It is obvious that the jumps of magnetostriction are caused by the change of the magnetic state of \MB\ at the phase transition. 

\begin{figure}[htbp]
	\centering
		\includegraphics[width=\stdPictureWidth]{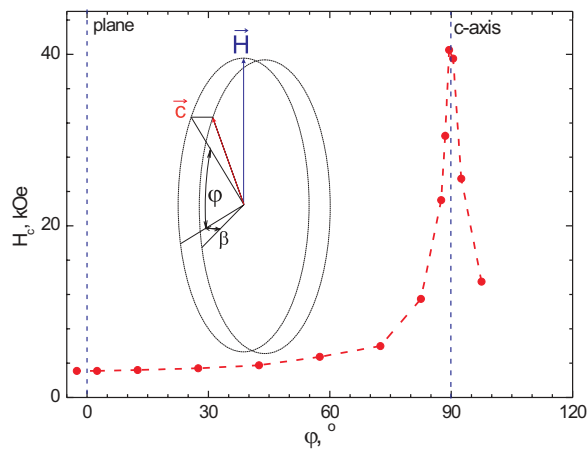}
	\caption{\add{Dependence of a critical field on the angle between the magnetic field and the basal plane of the crystal, $T=8$~K}}
	\label{fig:angle_dep}
\end{figure}
\begin{figure}[htbp]
	\centering
		\includegraphics[width=\stdPictureWidth,trim=0 0 0 0,clip]{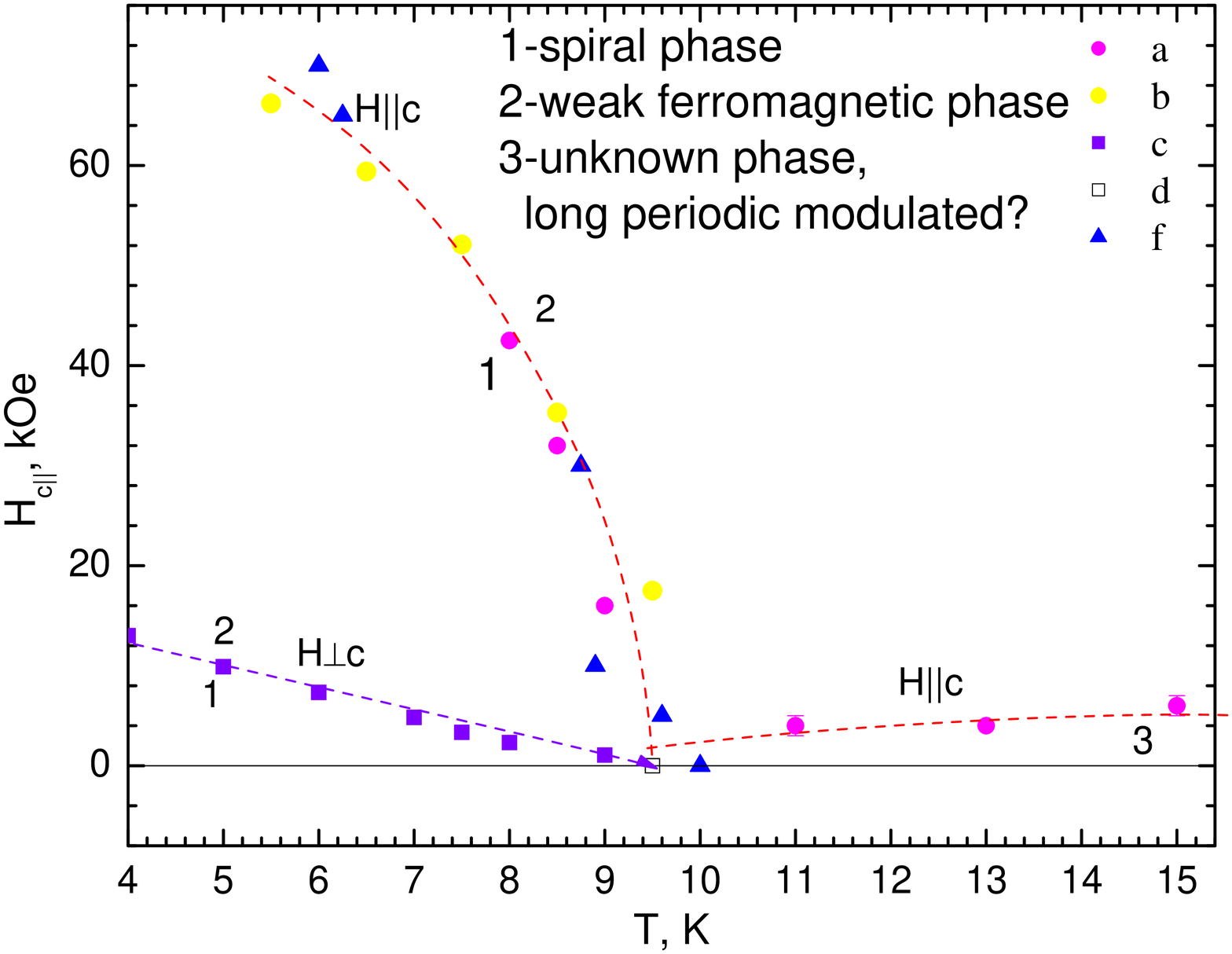}
	\caption{Phase diagram of \MB\ at $H\|c$. Data: a~--- transversal magnetization at $H\|c$, b~--- longitudinal magnetistriction at $H\|c$, c~--- longitudinal magnetization at $H\perp c$~\cite{Pankrats_2003}, d~--- heat capacity~\cite{JMMM_1999}, f~--- optical measurements at $H\|c$~\cite{Prl_2004}}
	\label{fig:phase_diag}
\end{figure}

Thus, the boundaries of magnetic phases of \MB\ magnetized along the tetragonal axis are established by resonant, magnetic and magnetostrictive measurements. We have measured also an angular dependence of the critical field at temperatures below $9.5$~K, this dependence has clearly defined maximum at $H\|c$\add{~(Fig.~\ref{fig:angle_dep})}. Because of this sharp dependence we show on the phase diagram~(Fig.~\ref{fig:phase_diag}) only the most authentic phase boundaries obtained by measurements of longitudinal magnetostriction and transversal magnetization at which the magnetic field was closest to the $c$-axis. \add{An estimate shows that at such angular dependence the inaccuracy of installation of samples at measurements in 2--3 degrees is capable to explain apparent spread of values of the critical field}. In addition to our data, the temperature of spontaneous phase transition measured by heat capacity~\cite{JMMM_1999} is marked, and the values of critical fields measured with the second optical harmonic generation ~\cite{Prl_2004} well agreed with our data are also presented. 
The resulted diagram is similar to the phase diagram \MB\ magnetized in a basal plane~\cite{Pankrats_2003}, but the critical fields for $H\|c$ are much higher. The state $1$ on the diagram corresponds to an incommensurate spiral phase, and the state $2$ is the field-induced weak ferromagnetic one. The nature of a state $3$ is unknown, we assume that it is also modulated, but the wave vector of modulation is much less than the resolution of neutron diffraction. 
The phase boundaries correspond to the first order transition and to the second one at temperatures, respectively, below and above ${T_{spon}=9.5}$~K.

\add{It is necessary to tell, that the existence of phase transition from incommensurate to commensurate state in a magnetic field laying in a basal plane is not surprised. In this case the field laying in a plane of a spiral deforms this structure, transforming it to fan, and then~--- to commensurate state. At the same time the phase transition in a magnetic field oriented along a 	wave vector of spiral structure seems at first sight surprising as energies of spiral and commensurate structures in a magnetic field at such magnetization are identical.}
We assume that the physical reason of incommensurate - commensurate phase transition at ${T<9.5}$~K and in $H\|c$  implies that the weak subsystem together with the strong subsystem plays the important role in a formation of spiral structure. And when the weak subsystem is saturated by a field along the tetragonal axis and its contribution to the formation of spiral structure changes.

The authors are greatly indebted to M.A~Popov and S.N.~Martynov for helpful discussions.

This work was supported by the Russian Foundation for Basic Research (grant RFBR 03-02-16701).

\end{document}